\documentclass[11pt]{article}
\usepackage{hyperref, url, graphicx, amsmath, amssymb, verbatim, booktabs, xcolor, fancyhdr, multirow, tikz, fontawesome}
\hypersetup{
    colorlinks=true,
    citecolor=VoatzBlue,
    linkcolor=VoatzBlue,
    urlcolor=VoatzBlue
    }
\definecolor{darkgreen}{rgb}{0,0.5,0}
\definecolor{LightBlue}{RGB}{18, 78, 126}
\definecolor{VoatzBlue}{RGB}{37, 62, 96}
\usepackage[margin=1in]{geometry}

\usetikzlibrary{shapes.geometric, arrows, calc}

\tikzstyle{votestep} = [rectangle, rounded corners, minimum width=3cm, minimum height=1cm, text centered, draw=black, text width=1in]
\tikzstyle{arrow} = [thick,->]

\newcommand{\lcm}{\operatorname{lcm}}
\newcommand{\ord}{\operatorname{ord}}
\newcommand{\ppmod}[1]{\mbox{\ $\left(\operatorname{mod}\ {#1}\right)$}}
\newcommand{\C}{\mathcal{C}}
\newcommand{\D}{\mathcal{D}}
\newcommand{\E}{\mathcal{E}}
\newcommand{\F}{\mathbb{F}}

\renewcommand{\S}{\mathcal{S}}
\newcommand{\Z}{\mathbb{Z}}

\begin{document}

\title{Overcoming Bottlenecks in Homomorphic Encryption for the 2024 Mexican Federal Election}
\author{Eric Landquist \quad Nimit Sawhney \quad Simer Sawhney \\ {\tt el@voatz.com} \quad {\tt ns@voatz.com} \quad {\tt ss@voatz.com}}
\maketitle

% \begin{abstract}
% \end{abstract}

\section{Introduction}
On June 2, 2024, Mexico held its federal elections. The majority of Mexican citizens voted in person at the polls in this historic election. For the first time though, Mexican citizens living outside their country were able to vote online via a web app, either on a personal device or using an electronic voting kiosk at one of 23 embassies and consulates in the U.S., Canada, and Europe. In total, 144,734 people voted outside of Mexico: 122,496 on a personal device and 22,238 in-person at a kiosk. Voting was open for remote voting from 8PM, May 18, 2024 to 6PM, June 2, 2024 and was open for in-person voting from 8AM-6PM on June 2, 2024. The voting and administrative software used for the expatriate component of the election was developed by the Boston-based software company Voatz and was run on Amazon Web Services (AWS).

In this article, we will describe the technical and cryptographic tools applied to secure the ex-patriate component of the election and to enable INE (Mexico's National Electoral Institute) to generate provable election results within minutes of the close of the election. The security of the voting platform was tested by an independent, third party team of experts chosen by INE. This team was given the source code of the voting app and documentation. Therefore, we designed the architecture assuming that an attacker had access to the source code, the ability to manipulate HTML and Javascript code in the browser on the frontend, and the ability to read and modify transmmissions between the frontend and backend.

Ballots were encrypted on the frontend via the Paillier cryptosystem, a homomorphic encryption scheme, therefore allowing votes to be tallied in real time while the ballots are encrypted, and without decrypting the current tally. To prove that votes were not tampered with, and also to maintain voter anonymity, the frontend generated a non-interactive zero-knowledge proof (ZKP) of the vote and the backend also applied a blind signature to the ballot. The voter received a ballot receipt code that could be used to verify the signature and the ZKP on the ballot after submission. This ensured that every ballot was recorded as cast and cast as intended. Section \ref{background} covers this cryptographic background.

One of the challenges in implementing homomorphic encryption was encoding ballot choice data to satisfy reporting requirements. Section \ref{encoding} describes our ballot encoding method. Mexican politics has a relatively complex coalition structure and the data needed to be encoded in order to properly report coalition results. A critical factor in tallying homomorphically encrypted ballots was to ensure that intermediate totals were not overwritten during periods of high volume voting. To this end, we used a high-throughput queueing service to ensure that every recorded ballot was tallied as recorded. This is described in Section \ref{flow}. This section also describes measures to prevent a voter from voting twice and a ballot from being submitted twice in a replay attack.

Another challenge was the need to decrypt each of the 144,734 ballots, each of which contained 2-4 encrypted contests, at the close of the election; 428,049 contest selections in all. This was necessary in order to verify the totals, process write-in candidates, and to generate reports for each embassy and consulate. A parallel approach was used to decrypt these ballots within minutes. This enabled relatively quick reporting of the results. Furthermore, independent observers, including Voatz personnel, could audit the results and verify that results were reported as tallied, ensuring the end-to-end integrity of the voting process. This administrative flow is described in Section \ref{admin}.

During the election, the backend verified the digital signatures and ZKP before recording a ballot. This was done in serial, leading to ballot submission times of 15-25 seconds, which was an acceptable, but not an ideal, duration. Section \ref{parallelize} gives results on how parallelization using AWS services improves ballot submission times.

While there were roughly 250,000 expatriate voters registered to vote remotely or at an in-person location outside of Mexico, we loaded the voter registration records of roughly 100 million voters (the entire eligible voting population of Mexico) onto our platform before the election. We stress tested the platform by benchmarking and fine tuning the performance of each architectural component to support such a volume. The success of this election demonstrates the feasibility of our approach to secure internet voting at scale. Section \ref{scaling} will also demonstrate how the solutions we present scale to elections on a national level.

\section{Background}
\label{background}

In this section, we describe the mathematical and cryptographic tools used in the Mexican elections, namely the Paillier cryptosystem, blind signatures, non-interactive zero-knowledge proofs, and Shamir secret sharing. The last subsection gives a brief description of blockchains.

\subsection{Paillier Cryptosystem}
\label{paillier}

Let $p$ and $q$ be distinct primes and $n = pq$. Let $\lambda = \lambda(n) = \lcm(p-1, q-1)$ and $g \in_{R}\Z_{n^2}^*$\footnote{$\Z_m$ denotes the ring of integers modulo $m$ and $\Z_m^*$ denotes the multiplicative group $\{a\in\Z_m \vert \gcd(a, m)=1 \}$.} such that $n|\ord(g)$. For any $\alpha, \beta\in\Z_n^*$, $g = (\alpha n+1)\beta^n \ppmod{n^2}$ satisfies this criterion. In particular, we can let $g=n+1$. Next, let $L: \Z_{n^2}^* \to \Z_n$ be given by $L(u) = \frac{u-1}{n}$, where $u\equiv 1\ppmod{n}$, and let $\mu = \left(L\left(g^{\lambda}\ppmod{n^2}\right)\right)^{-1}\ppmod{n}$. A user's Paillier public key is the pair $(n, g)$ and the corresponding private key is the pair $(\lambda, \mu)$.

Encryption is a randomized function $\E: \Z_n \to \Z_n^*$ defined as follows. To encrypt $m\in \Z_n$, randomly choose $x\in_R \Z_n^*$\footnote{Throughout this document, $\in_R$ signifies that the element was chosen from the set at random, using a cryptographically secure random number generator.}, and compute
$$c = \E(m) = g^mx^n \ppmod{n^2} \enspace .$$
Decryption, $\D:\Z_{n^2}^* \to \Z_n$, is defined
$$m = \D(c) = L\left(c^{\lambda}\ppmod{n^2}\right) \mu \ppmod{n} \enspace .$$

The Paillier cryptosystem relies on the presumed intractibility of the {\em composite residuosity problem}, i.e, the problem of computing $n$th roots modulo $n^2$, and the {\em decisional} composite residuosity problem, the problem of deducing whether or not an element $z\in\Z_{n^2}^*$ is an $n$th root residue, for its security. In other words, given an element $z\in\Z_{n^2}^*$, it is believed to be computationally difficult to determine $x\in\Z_{n^2}^*$ such that $x^n \equiv z \ppmod{n^2}$ \cite{paillier}. The integer factorization problem reduces to these two problems, so the security of $n$ as used for an RSA modulus is a reasonable approximation for the security of Paillier parameters made using the factors of $n$.

The Paillier cryptosystem is partially (additively) homomorphic. If $m_1, m_2 \in \Z_n$ and $x_1, x_2 \in_R \Z_n^*$ then:
$$\E(m_1)\cdot\E(m_2) \equiv \left(g^{m_1}x_1^n \right)\left(g^{m_2}x_2^n \right) \equiv g^{m_1+m_2}(x_1x_2)^n \equiv \E(m_1+m_2) \ppmod{n^2}\enspace .$$
(In our case, a plaintext message was a vector that encoded a voter's ballot choices. Details on this encoding are in Section \ref{encoding}.) Therefore, if $\{m_1, \ldots, m_k\}$ is a set of encoded ballots, then a running tally of the ballots can remain encrypted by taking the modular product of the encrypted ballots:
$$\prod_{i=1}^{k}\E(m_i) \ppmod{n^2} = \E\left(\sum_{i=1}^km_i\right) \enspace .$$

In our implementation, we utilized the package {\tt java.security} to generate a pair of cryptographically secure 1536-bit primes so that the modulus $n$ was 3072 bits, equivalent to 128-bit symmetric key security~\cite{nist}.

An attacker can alter an encrypted ballot, $\E(m)$, after submission by picking any $k\in\Z$ and computing $\E(m)^k \ppmod{{n^2}} = \E(km)$. If $k>0$, this action changes the ballot to cast $k$ votes at once. If $k=0$, this erases the ballot. If $k<0$, then this effectively deletes multiple ballots. We can avoid this attack by computing a {\em zero-knowledge proof} of the ballot.

\subsection{Zero-Knowledge Proofs}
\label{zkp}

A {\em zero-knowledge proof} (ZKP) is a cryptographic protocol that gives a computationally verifiable proof that someone (such as a voter) possesses knowledge of some information (such as his vote) without revealing the information, whether in whole or in part. The first ZKPs were interactive, requiring a series of challenges and responses between the prover (who possesses some information) and the verifier (who does not have that information). However, such interaction risks compromising the identity of the voter and is infeasible to perform after a ballot has been submitted and recorded. Therefore, we utilized a {\em non-interactive} ZKP to enable ballot verification after submission and recording.

Baudron et al.\ \cite{b} adapted the (interactive) ZKP of Guillou and Quisquater \cite{gq} and Okamoto \cite{o} to the Paillier cryptosystem. In Step 3 below, we applied the trick of Fiat and Shamir \cite{fs} to make the ZKP non-interactive. We also made a minor, but necessary, simplification, without which, we would need the private Paillier key in order to create a valid proof. The following protocol is the ZKP that the prover (which in our case is the voter, or, in practice, the web app frontend) knows both the vote $m$ and the randomizer $x$ used to encrypt $m$. The final step shows how the verifier (the backend or any election auditor) verifies the proof.

\begin{enumerate}
    \item The voter votes $m$, and the frontend generates the randomizer $x$ and computes $c=\E(m) = g^mx^n\ppmod{n^2}$.
    \item The frontend first chooses $r \in_R \Z_n$ and $s \in_R \Z_n^*$, and computes the {\em commitment} $u = g^rs^n \ppmod {n^2}$.
    \item The frontend computes the (non-interactive) {\em challenge} $e = H(u) \ppmod{2^{48}}$, where $H$ converts $u$ to its base 64 string representation, computes the SHA-256 digest in base 64, then converts that result to the corresponding integer.
    \item We require $r > em$, so we check that this inequality is satisfied.
    \item The frontend computes $v= r - em$ and $w = sx^{-e} \ppmod{n}$.
    \item The frontend sends the backend the encrypted ballot $\E(m)$ and the ZKP $(u, e, v, w)$.
    \item The backend checks if $u = g^vc^ew^n \ppmod{n^2}$ to verify the ZKP.
\end{enumerate}
The commitment $u$ is a standard binding and hiding commitment scheme. The prover is bound to $r$ and $s$ since $g^r$ and $s^n$ are each unique modulo $n^2$, so it is computationally infeasible to find another $r_0 \in_R \Z_n$ and $s_0 \in_R \Z_n^*$ such that $u \equiv g^{r_0}s_0^n\ppmod{n^2}$. Further, an adversary cannot determine any information about $r$ and $s$ from $u$, since they are chosen at random and $u$ is computationally indistinguishable from random elements of $\Z_{n^2}^*$ under the decisional composite residuosity assumption; $r$ and $s$ are hidden. If the encrypted ballot, $c=\E(m)$, is modified to some value $C$ in transit or in storage, then the check in the last step will fail; the value of $g^vC^ew^n \ppmod{n^2}$ will not equal the commitment $u$. Moreover, it is computationally infeasible for an attacker to modify the ZKP into a valid ZKP if the ballot is modified.

The requirement that $r>em$ ensures that $0 < v < n$, a condition necessary for the ZKP to be valid. Notice that in the verification of the ZKP, the verifier computes $g^v\ppmod{n^2}$, so in order to remove this restriction, the prover would need to compute $v = r-em \ppmod{|\Z_{n^2}^*|}$.\footnote{The {\em order} of a group, $G$, is the number of elements in that group and is denoted $|G|$.} However, $|\Z_{n^2}^*| = \varphi(n^2) = n(p-1)(q-1)$, so the prover (i.e., frontend) would need to know the factorization of $n$ (or, equivalently, the Paillier private key) in order to compute $v$. Now the requirement that $r > em$ restricts the length of a message that we can encrypt in order to ``make room for" the ZKP. Since $r\in\Z_n$, $r$ will be 3072 bits in general and $e$ will be 48 bits. Therefore, in our application we needed to restrict encoded ballots to less than 3024 bits in length. In fact, we restricted ballots to 3001 bits to make it very likely for $r>em$ to always hold.

Before ballot submission, however, the frontend requested a blind signature on the ballot.

\subsection{Blind Signatures}
\label{blind-sig}

Blind signatures were introduced by Chaum in \cite{chaum} for anonymous electronic cash transactions. The same principles apply to electronic ballot submission to maintain voter anonymity for ballot submission.

Let $p,q$ be cryptographically secure primes and $n=pq$. Let $e = 2^{16} +1 = 65537$ be the election's (public) signature verification key\footnote{Any $e\in\Z_n^*$ can be used for the public signing key, but the prime 65537 is typically used for efficient signature verification.} and $d = e^{-1} \ppmod{\varphi(n)}$ be the corresponding (private) signing key; $\varphi(n) = (p-1)(q-1)$. For the frontend to obtain a blind signature on a ballot $c = \E(m)$, the following procedure is followed. In this protocol, the function $H$ is the same SHA-256 hash function as used in the ZKP protocol.

\begin{enumerate}
    \item The frontend generates a {\em mask} (also called a {\em blinding factor}) $r \in_R \Z_n^*$.
    \item The frontend computes $b = H(\E(m)\|\E(w))\cdot r^e \ppmod{n}$, where $\|$ denotes concatenation and $w$ is a write-in candidate.
    \item The frontend sends $b$ over an authenticated connection to the backend.
    \item The backend verifies that the voter has not voted and is a valid voter.
    \item The backend returns $B = b^d \ppmod{n}$, its signature on the blinded ballot, to the frontend.
    \item Since $B \equiv b^d \equiv H(\E(m)\|\E(w))^d\cdot r \ppmod{n}$, the frontend computes $s = r^{-1}B \ppmod{n} \equiv H(\E(m)\|\E(w))^d \ppmod{n}$, which removes the blinding factor and results in the election's digital signature on the hashed ballot.
    \item The frontend verifies the election's signature by verifying that $s^e \equiv H(\E(m)\|\E(w)) \ppmod{n}$.
    \item The frontend submits the ballot package $(\E(m), \E(w), s)$ to the backend.
    \item The backend verifies its signature by checking that $s^e \equiv H(\E(m)\|\E(w)) \ppmod{n}$.
\end{enumerate}

In our implementation, we utilized {\tt java.security} to generate another pair of cryptographically secure 1536-bit primes, so that the 3072-bit modulus for the signing key would differ from the Paillier modulus. The election encryption and signing keys were generated before the start of the election in a key generating ceremony overseen by INE. While it was necessary to persist the private signing key in a table in DynamoDB, a serverless NoSQL database service, in order to authorize ballot submissions, the ballot decryption key had to be destroyed. This enabled voter anonymity and prevented the release of intermediate ballot tallies before the end of the election. In order to recover the key, we applied a threshold (secret sharing) scheme to split up the election decryption key.

\subsection{Shamir Secret Sharing}
\label{threshold}

INE required the election decryption key to be split among five election officials in a way that any three of them could reassemble the key. A different {\em shard} of this key was distributed to each official on a thumb drive. Then the actual decryption key was destroyed. After the close of the election, this key was reassembled and the election results were decrypted. We used Shamir's threshold scheme, which is based on the fact that there is a unique $t-1$-degree polynomial passing through $t$ distinct points \cite{shamir}.

Shamir's scheme works as follows. Suppose that there is an $M$-person committee such that $t\le M$ committee members are required in order to obtain the decryption key, $\lambda$. (So in our case, $M=5$ and $t=3$.) Let $r\in\Z$ such that $r>n$ and $r$ is prime.\footnote{$n$ is the Paillier modulus.} For $2\le i<t$, generate $a_i\in_R\F_r$, and let
$$f(x) = \lambda + \mu x + \sum_{i=2}^{t-1}a_ix^i \in \F_r[x] \enspace .\footnote{$\F_r = \{0, 1, \ldots, r-1\}$ is the finite field of integers modulo $r$ and $\F_r[x]$ is the ring of polynomials with coefficients in $\F_r$.}$$
For each committee member, choose distinct $x_i\in_R\F_r$ and compute $y_i = f(x_i)\ppmod{r}$. Committee member $i$ receives the point $(x_i, y_i)$. To reconstruct the key, we compute the $(t-1)$-degree Lagrange interpolating polynomial, $P(x)$, through any $t$ points:
$$P(x) = \sum_{i=0}^{t-1}y_i\prod_{\substack{0\le j < t\\ j\ne i}}(x-x_j)(x_i-x_j)^{-1}\ppmod{r} \enspace .$$
This is called an $M$-$t$ {\em Shamir threshold scheme}.

\subsection{Blockchains}
\label{blockchain}
Our architecture utilized a {\em blockchain} to record anonymous ballot IDs for audit purposes; details will be described in Section \ref{flow}. A blockchain is a linked list of {\em blocks} of data that is replicated on multiple nodes. A cryptographically secure hash (e.g., SHA-256) of one block is contained in the subsequent block to establish the link between the two blocks. The first block of a blockchain is called its {\em genesis block}. A {\em consensus mechanism} describes the method in which the nodes agree on whether or not a proposed block satisfies the conditions necessary to be added to the blockchain.

A blockchain has a natural application to elections. Data recorded on a blockchain is immutable and time stamped, so election data can be preserved and audited by the public. Voatz uses a blockchain, namely Hyperledger Fabric, as a key component of our platform for this very purpose. In all other elections, we have stored anonymous ballot choices directly on a blockchain, but for the Mexican election, we only stored anonymous ballot IDs on the blockchain. This established an independent accounting for each ballot that was cast.

In the next section, we show how we encoded the ballot in order to enable tallying and reporting.

\section{Ballot Encoding}
\label{encoding}

In this section, we describe the ballot encoding method used and the challenges that the coalition structure posed.

For the Mexican elections, each contest on a ballot was encrypted separately. Each ballot had contests for President and Senate and some states had contests for Governor and/or another local office. So each state had between two and four contests. Reports needed to be run for each state and modality, so there were 150 separate contest totals, 75 each for in-person voting and remote voting, with 32 President totals, 32 Senate totals, and 11 state-specific contest totals.

The choices in each contest were encoded as components of a vector. A final component encoded the number ``1" so that when the ballots were tallied, the last component encoded the ballot total as a check. The vector was initially encoded as a binary string, read right to left, for a natural conversion into an integer for encryption. There were 234,117 people registered to vote remotely or in person, so choice totals and ballot totals would be well under $2^{20}$ for each contest. Therefore, 20 bits was allocated to each vector component -- enough space to fit tallies up to $2^{20} -1$.

By way of example, suppose a contest had four candidates and there were less than 32 voters, then we could encode each choice using $\log_2(32) = 5$ bits. In this case, a vote for Candidate \#2 would be encoded:
$$\langle1, 0, 0, 1, 0\rangle \to \langle1, 00000, 00000, 00001, 00000\rangle \to 1 00000 00000 00001 00000_2 = 2^{20} + 2^5 = 1048608\enspace .$$

As noted earlier, we restricted ballots to 3001 bits; this enabled the encoding of up to 150 choices, with an extra bit for the total number of encoded ballots, i.e., ``1."

\begin{figure}[ht]
    \centering
    \includegraphics[width=0.6\textwidth]{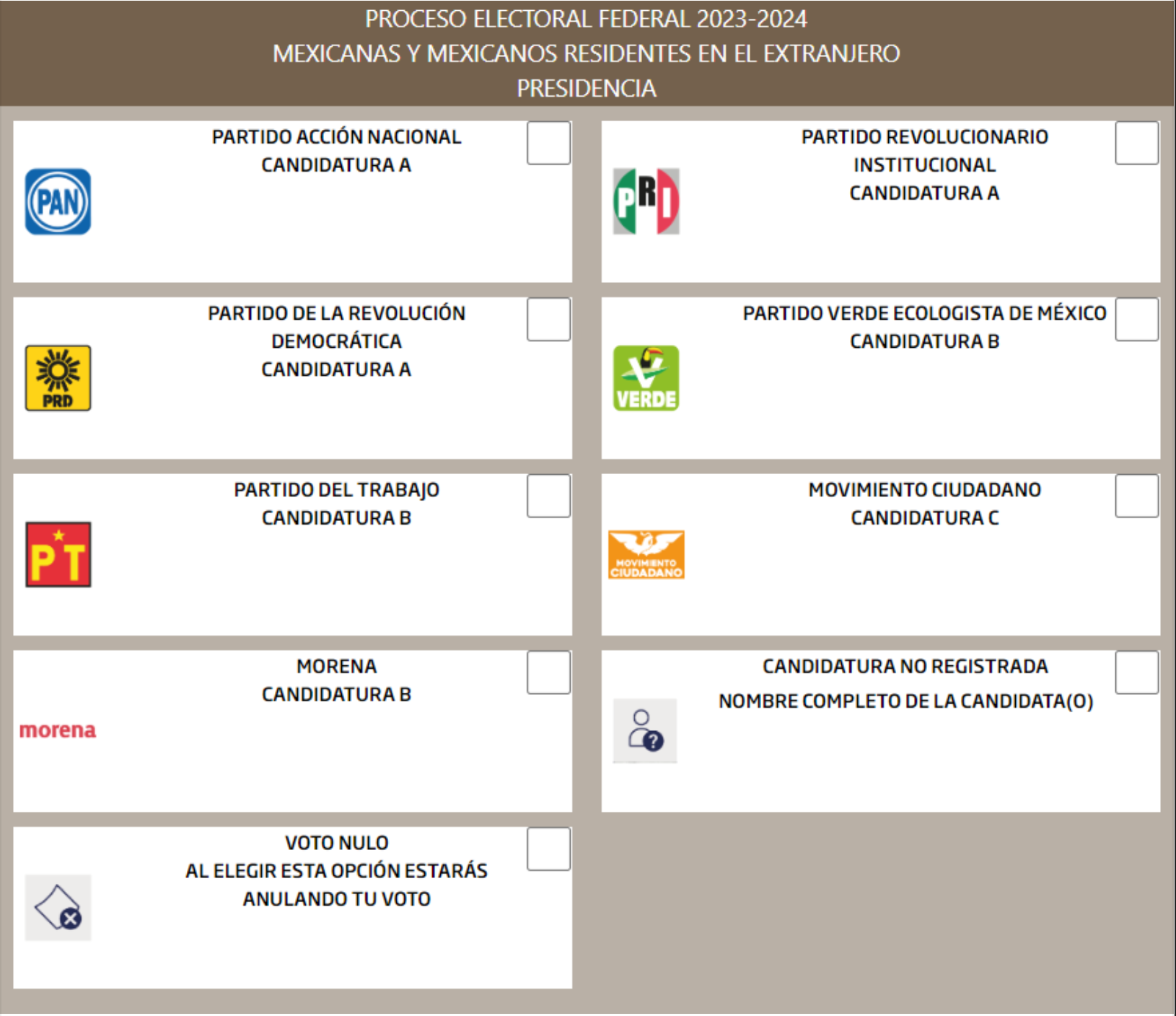}
    \caption{Test Presidential Ballot with Coalition Structure}
    \label{fig:coalitions}
\end{figure}

Mexican politics is structured by coalitions; for many offices, a voter votes for a particular candidate by way of voting for a party or parties that endorse that candidate. In the race for President, for example, there were three candidates on the ballot, with options to write in a candidate or to select ``No Vote." (In the web app, at least one option had to be selected.) See Figure \ref{fig:coalitions} for a test ballot for President with three test (placeholder) candidates and seven parties.
Here, ``Candidatura A" was endorsed by a coalition of three different political parties (PAN, PRI, and PRD), ``Candidatura B" was endorsed by a second coalition of three other parties (Verde, PT, and Morena), and a third candidate was endorsed by a seventh party (MC). In this format, a voter could choose to select any number of parties in that contest, provided that all parties selected were in the same coalition. So, for example, a voter could select the two parties PAN and PRI because both backed ``Candidatura A," but could not select the two parties PAN and Morena because those two parties are in different coalitions.

In the case that a voter selected a write-in, the name of the write-in candidate was Paillier encrypted and stored in a DynamoDB table with the rest of the encrypted ballot package. The contest tally therefore only recorded the total number of ballots that had a write-in for that contest. Individual ballots had to be decrypted in order to report on the actual names of the candidates that had been written in.

Reports needed tallies of each possible combination of selections for each contest. So in the example of the race for President, we encoded $2^3-1 = 7$ choices for each 3-party coalition and one each for the seventh party, write-in, and no vote, a total of 17 choices. In general, a $k$-party coalition needed $2^k-1$ choices in the contest vector to represent each possible valid selection within that coalition. To determine which component of the contest vector represented a particular subset of parties in a coalition, we used a binary scheme. All single-party selections were encoded in the first components of the contest vector, followed by coalition selections. Parties were ordered in a coalition based on their prescribed order on the ballot and numbered $0$ to $k-1$. If a coalition consisted of the parties $\C = \{p_0, ..., p_{k-1}\}$ and a voter selected a subset $\S\subseteq\C$ of these parties, then the coalition choice is numbered $\left(\sum_{p_i\in\S}2^{i}\right) - 2$. Since we did not encode single-party choices here, the smallest coalition choice number corresponds to the subset $\S=\{p_0, p_1\}$, yielding the coalition choice number $2^0 + 2^1 - 2 = 1$. So in the case of the contest for President, we encoded the first nine vector components for single-party selections, write-ins, and no-votes. The next five were reserved for the first coalition, and the last five for the second coalition; 19 vector components in total.

One particular challenge was the case of the race for Governor in the state of Chiapas. In this contest, there were 13 political parties on the ballot, with options for write-ins and no-votes. There were two coalitions, one with 9 parties, and another with 3. Therefore, there were a total of $2^9-1 + 2^3-1 + 3 = 511$ possible ballot selections for that contest, far more than the space allowed by the Paillier modulus and ZKP. For that contest, we divided the choice vector into four separate vectors and encrypted each, computing a separate ZKP for each of the four sub-vectors. The digital signature was applied to the entire contest, however. The encrypted contest, encrypted total, and ZKPs were stored as virtual arrays. This approach is effective for any contest in which the number of options (or candidates) in an election exceeds the capacity of a single choice vector.

This section and the previous section described the encoding and cryptographic tools that we used. In the next two sections, we will zoom out to describe the overall voting flow and administrative flow, with more details on the AWS platform services that we used.

\section{Voting Flow and Ballot Queueing}
\label{flow}

In this section, we describe the overall flow of the voting process. Of particular note is the use of AWS SQS (Simple Queue Service) to ensure that the ballot tallies were accurate. We also describe measures to prevent a ballot from being submitted twice and also to maintain voter anonymity.

From May 4-15, 2024, voters were invited to onboard into the voting app and get acquainted with the interface before the election opened for remote voters on May 18th. During this orientation period, those voting remotely could verify their identity, create a password, and set up 2-factor authentication for subsequent log in. After the close of the orientation period, voters were able to onboard during the live voting period.

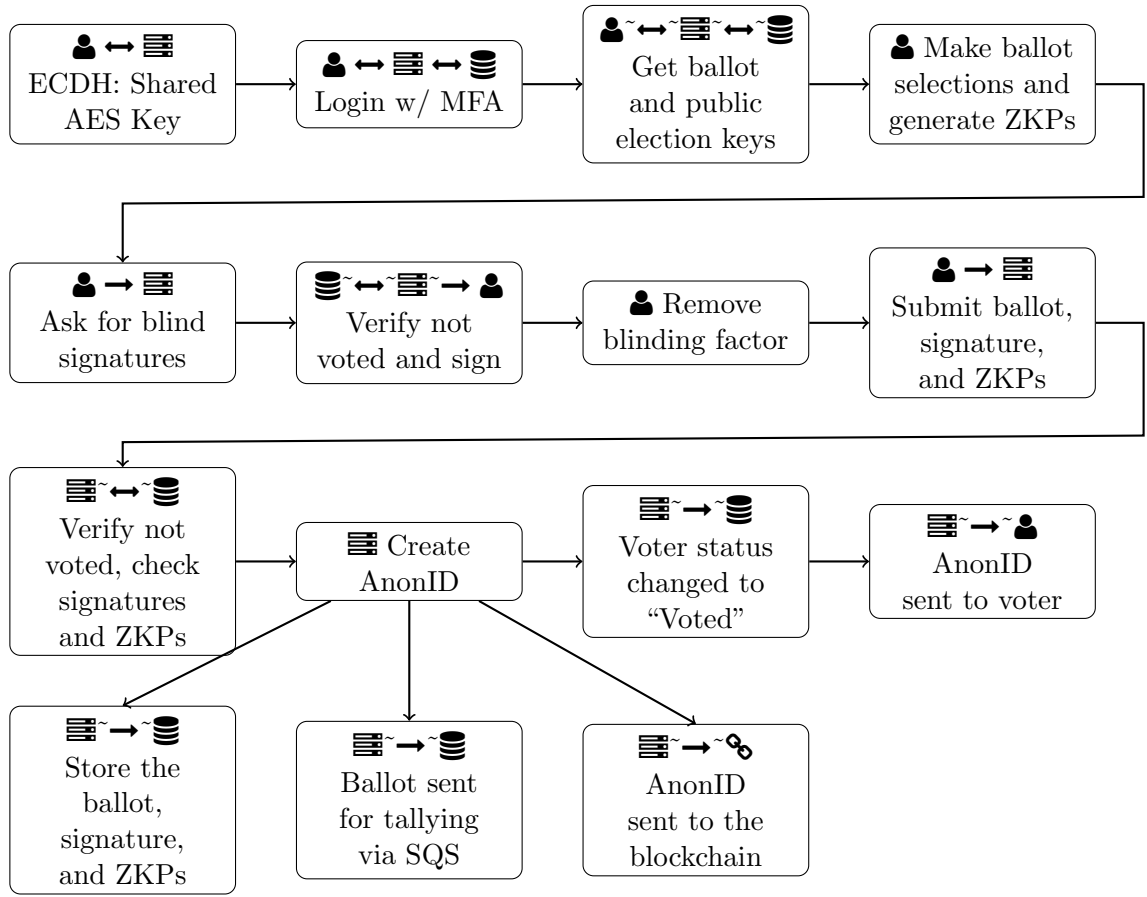
\begin{figure}[ht]
    \centering
\begin{tikzpicture}[node distance=1.5in]
% Nodes
\node (start) [votestep] {\faUser\ \faArrowsH\ \faServer\ ECDH: Shared AES Key};
\node (step1) [votestep, right of=start] {\faUser\ \faArrowsH\ \faServer\ \faArrowsH\ \faDatabase\ Login w/ MFA};
\node (step2) [votestep, right of=step1] {\faUser\~\faArrowsH\~\faServer\~\faArrowsH\~\faDatabase\  Get ballot and public election keys};
\node (step3) [votestep, right of=step2] {\faUser\ Make ballot selections and generate ZKPs};
\node (step4) [votestep, below of=start, yshift=0.25in] {\faUser\ \faLongArrowRight\ \faServer\ Ask~for blind signatures};
\coordinate (midpoint) at ($(start)!.5!(step4)$);
\node (step5) [votestep, right of=step4] {\faDatabase\~\faArrowsH\~\faServer\~\faLongArrowRight~\faUser\ Verify not voted and sign};
\node (step6) [votestep, right of=step5] {\faUser\ Remove blinding factor};
\node (step7) [votestep, right of=step6] {\faUser\ \faLongArrowRight\ \faServer\ Submit ballot, signature, and ZKPs};
\node (step8) [votestep, below of=step4, yshift=0.25in] {\faServer\~\faArrowsH\~\faDatabase\ Verify not voted, check signatures and ZKPs};
\coordinate (midpoint2) at ($(step4)!.5!(step8)$);
\node (step9) [votestep, right of=step8] {\faServer\ Create AnonID};
\node (step10) [votestep, right of=step9] {\faServer\~\faLongArrowRight\~\faDatabase\ Voter~status changed to ``Voted"};
\node (end) [votestep, right of=step10] {\faServer\~\faLongArrowRight\~\faUser\ AnonID sent to voter};
\node (ballot) [votestep, below of=step8, yshift=0.25in] {\faServer\~\faLongArrowRight\~\faDatabase\ Store~the ballot, signature, and ZKPs};
\node (tally) [votestep, below of=step9, yshift=0.25in] {\faServer\~\faLongArrowRight\~\faDatabase\ Ballot sent for tallying via SQS};
\node (chain) [votestep, below of=step10, yshift=0.25in] {\faServer\~\faLongArrowRight\~\faChain\ AnonID sent to the blockchain};

% Arrows
\draw [arrow] (start) -- (step1);
\draw [arrow] (step1) -- (step2);
\draw [arrow] (step2) -- (step3);
\draw [arrow] (step3.east) -- ++(0.25in,0) -- ++(0,-1.5) -- (midpoint) -- (step4.north);
\draw [arrow] (step4) -- (step5);
\draw [arrow] (step5) -- (step6);
\draw [arrow] (step6) -- (step7);
\draw [arrow] (step7.east) -- ++(0.25in,0) -- ++(0,-1.5) -- (midpoint2) -- (step8.north);
\draw [arrow] (step8) -- (step9);
\draw [arrow] (step9) -- (step10);
\draw [arrow] (step10) -- (end);

\draw [arrow] (step9) -- (ballot.north);
\draw [arrow] (step9) -- (tally.north);
\draw [arrow] (step9) -- (chain.north);
\end{tikzpicture}
\caption{Voting app flow.}
\label{fig:flow}
\end{figure}

The diagram in Figure \ref{fig:flow} illustrates the flow of the voting app. Each stage indicates actions by the voter or frontend (\faUser), the backend (\faServer), a database (\faDatabase), and the blockchain (\faChain), along with communication between these components (\faArrowsH\ and \faLongArrowRight\ ).

When a voter visited the voting website, the backend and frontend engaged in an ECDH (Elliptic Curve Diffie-Hellman Key Exchange) handshake using NIST Curve P-256 to establish an ephemeral shared secret state, which was used to create a shared 256-bit AES key to encrypt all payload data. (This is an application layer encryption over and above TLS.) This shared state, and hence the secret key, was changed deterministically after each transmission between the frontend and backend. So if any transmission between the frontend and backend were intercepted, it could not be resent; decryption would fail. In particular, this is one layer of defense preventing a ballot from being submitted twice.

Upon log in during the live voting period, the backend sent the voter the ballot and the election's public encryption and signing keys. For each contest, the frontend encrypted the contest selections and generated a ZKP of the vote. Upon ballot submission, the frontend obtained a blind signature on the ballot, verified the signature, removed the blinding factor, and submitted the ballot package consisting of the encrypted contests, the ZKPs, and the digital signatures. The backend checked the voter's voting status, verified the signature and the ZKPs, and checked that the correct ballot was submitted. If all checks passed the backend hashed the encrypted ballot to generate an anonymous ballot confirmation ID, called the ``AnonID." (We stress that there was no computational link between the AnonID and the voter's identity and that the voter's identity was never persisted with either his ballot or the AnonID.) Each encrypted contest was persisted in a DynamoDB table, keyed by the AnonID. The AnonID was also recorded on the election's blockchain to provide an independent and immutable accounting for that ballot. Finally, the backend sent each encrypted contest to AWS SQS for tallying and returned the AnonID to the frontend. (The voter portal dashboard contained a link allowing voters to verify the ZKP and blind signature on any AnonID, thus checking their ballot for tampering.) Meanwhile (i.e., concurrently with the storage and tallying of the encrypted ballot), the backend sent a request to a MySQL database to mark that voter as having voted. We note that this EC2 instance was physically and logically distinct from the NoSQL (DynamoDB) tables that stored the encrypted ballots to ensure that voters could not be paired with their ballots. So ballot submission was processed in the same way as the ``double envelope" system used to preserve voter privacy and anonymity with physical mail-in ballots.

Another safeguard to prevent a single voter from voting multiple times were aggressive measures, such as TTL (time to live) based in-memory caching, to prevent any user from having multiple open simultaneous sessions. This prevented voters from submitting a second ballot while the first was in process.

Now each of the submitted ballots entered an SQS queue for a specific contest in a specific state or district, for the given voting modality (remote or in-person). The queues operated on a first-in-first-out (FIFO) basis, with deduplication checks to ensure that each ballot is tallied exactly once. To enter a queue, a contest selection had to pass a deduplication check: the concatenation of the ballot's AnonID and contest ID had to be unique. This is another layer of defense preventing a ballot from being submitted twice. The following then took place when an encrypted contest reached the front of the queue. First, the digital signature was checked, preventing tampering of the totals via insertion attacks. Then, the current encrypted total for that contest, state, and modality was retrieved from the totals table. The new encrypted total was computed and updated, with the ballot count getting incremented. In the event that a certain ballot was the first ballot for that contest, state, and modality, then that encrypted contest was stored directly in the totals table and the ballot total was set to 1.)

A service that processes ballots in a queue, like SQS, was necessary in order to prevent ballots from getting overwritten in the totals. Other possible solutions are discussed below. Each API call from the frontend to the backend is processed by an AWS Lambda function, a serverless computing service. Lambda functions scale automatically, so that multiple invocations of a single Lambda function can run concurrently if so triggered by a large volume of incoming traffic. Without SQS, the following race condition would be possible during periods of high voting volume. Suppose that two voters, Voters A and B, registered in the same state, cast their ballots via the same modality at about the same time. The ballot submission Lambda function is triggered by the API call for the ballot from Voter A. This Lambda invocation reads the totals DynamoDB table for each contest in Voter A's ballot. While the Lambda function is computing the (homomorphic) product of the current totals and the Voter A's contests, Voter B's ballot arrives in the backend, invoking a second instance of the ballot submission Lambda function. This invocation reads the totals DynamoDB table for each contest in Voter B's ballot -- data that is precisely the same as what was read by Voter A's Lambda invocation. While this invocation is computing the new total for each contest in Voter B's ballot, Voter A's invocation updates the DynamoDB table with the new total that tallies Voter A's ballot. Shortly thereafter, Voter B's Lambda invocation updates the DynamoDB table with the new total that tallies Voter B's ballot, overwriting the update made by Voter A's Lambda invocation. Voter A's ballot, while still stored in a separate ballot table and marked as submitted, is effectively erased from the totals.

Making a separate Lambda function that solely performs updates of Paillier-encrypted totals is not sufficient. From the 144,734 ballot submissions, there were 428,040 individual encrypted contests, each of which invoked the totals update Lambda function, with an average duration of 174ms. The SQS service routed ballot choices into one of 150 different queues, one for each separate tally. During the live election, there were up to 10 concurrent invocations of the Lambda function that updated homomorphic totals. Other than seven spikes in which queued ballots waited between 10 and 54 seconds to be processed, ballots did not have to wait more than 1 second in a queue. The maximum length of any one queue was 3 ballots, with 9 separate instances in which ballots had to wait in a queue to be processed.

Therefore, when using a homomorphic tally of ballots, it is crucial to process ballots in a FIFO manner in order to eliminate any possibility of a ``read-read-update-update" race condition that would overwrite the current running tally. There are some approaches other than SQS for homomorphic tallying that are worth noting. First, a blockchain provides an effective decentralized approach, given its properties described in Section \ref{blockchain}. In this scenario, the genesis block would contain a list of all contests in the election, assigning the multiplicative identity ``1" as the initial (homomorphic) total for each contest. Each block would contain homomorphic tallies of the ballots. During the election window, nodes processing the blockchain would compute sub-tallies of contests from each ballot pending inclusion in the current total, and update tallies accordingly for each new block, with consensus from all other nodes in the network that validate each proposed update to the blockchain. This is a highly resilient and secure solution since multiple copies of the results would be distributed geographically. A second alternate approach would store totals in a relational database, since relational databases lock records during table updates. One performance downside is that relational databases can be comparatively slower than non-relational databases. A third option was considered that used DynamoDB with a technique called optimistic locking, which prevents updates from getting overwritten. With this approach, however, an error is thrown if an update is attempted while another update is in progress. This approach was not pursued because this scenario was expected during periods of high voting volume and theoretically, a ballot would continue to be rejected for an indeterminate length of time. The behavior of the application would have been much less predictable with this kind of architecture. Also, in a philosophical sense, such an update should not be an error, but a situation in which the ballot should simply wait in line until it's that ballot's turn to be counted.

Since the total number of ballots cast for each contest, state, and modality were encrypted with the total, as well as persisted in the clear in the DynamoDB totals table, these figures could be compared with each other upon decryption and checked in real time against the total number of voters who have voted, as displayed in the administration portal, and against the total number of AnonIDs recorded in the blockchain. These totals were all equal, proving that the chosen architectural components ensured correctness of the application.

\section{Administrative Flow}
\label{admin}

In this section, we describe the administrative flow and tasks as they relate to the cryptographic protocols needed for voting and reporting. A key component of this process is the parallelization of individual ballot decryption to enable complete results to be generated and verified efficiently.

Before the live election, INE held two week-long end-to-end simulations to test every aspect of the administrative and voting process. Both simulations and the live elections followed the following administrative procedure. The Paillier and digital signature keys were generated in a key generation ceremony on an air gapped system with no network connectivity. The private Paillier decryption key was split into 5 shards via a $5$-$3$ Shamir threshold scheme, with each of five committee members receiving a thumb drive and a back-up drive with a shard. This private key was then destroyed and the public encryption and digital signature keys were uploaded to a DynamoDB table.

After the election closed, the Paillier decryption key was reassembled, the remote and in-person ballots and totals were downloaded from their respective DynamoDB tables. Totals were decrypted immediately, with reports run on overall totals. Complete reports with all write-ins, as well as reports specific to each in-person voting location, required officials to decrypt the individual ballots. With 428,040 individual contests to decrypt (364,904 remote and 63,136 in-person contest choices), this would be a time-consuming endeavor, so we parallelized ballot decryption, dividing the in-person contests into 4 separate files and the remote contests into 32 separate files. After decryption, the decrypted ballots were combined into a single file and final reports were generated. Decryption of the in-person contests took about 3.2 minutes and the decryption of the remote contests took about 4.6 minutes of clock time.

\section{Parallelizing Ballot Submission}
\label{parallelize}

In this section, we give results from fully parallelizing the ballot submission process. In the Mexican election, we parallelized the computations to update the ballot totals via SQS. Here, we used AWS Step Functions to parallelize the process of obtaining and verifying digital signatures and verifying ZKPs.

There are two separate Lambda functions that are invoked in the ballot submission process. The first obtains a blind digital signature on a ballot. The second verifies the digital signature, the ZKP, and processes the ballot. In the Mexican election, blind signatures were made in serial, and the blind signatures and ZKPs were both verified in serial before the ballot was processed. These two Lambda functions ran for an average of 0.905 seconds and 10.974 seconds, respectively, for a total of 11.879 seconds of backend processing time. Note that this does not include latency time of the two API calls nor the time for the frontend to verify the blind signatures for itself. Obviously, this time would have been longer for ballots with three or four contests. So depending on ballot length, latency, and the processor speed of the voter's device, the overall ballot submission time was typically in the range of 15-25 seconds. (For this reason, there was a visual wait indicator -- a spinner -- and other safeguards to prevent a user from going back and attempting to resubmit a ballot.) Now lambda functions run on a single core, so parallelization requires other computing services.

As noted, we parallelized the ballot submission process using AWS Step Functions. Step Functions allow one to create an automated workflow by combining various computing services. Parallelizing a Lambda function using Step Functions is relatively straightforward, using the {\tt Map} and {\tt Parallel} workflows. The {\tt Parallel} workflow allows two separate tasks to run in parallel, such as two distinct Lambda functions, for example. The {\tt Map} workflow allows one to run multiple invocations of a certain task in parallel. In our implementation, the Lambda function to obtain blind signatures was parallelized via the {\tt Map} workflow, running one Lambda instance for each contest in parallel. The second stage of ballot submission required three new Lambda functions: one to verify the blind signatures, one to verify the ZKP, and one to process the results of the various verifications.

In this case, the ballot submission Lambda checks and updates the voter's voting status, creates the AnonID, calls the Step Function, then returns the AnonID to the frontend. (The ballot submission Lambda need not wait for the verifications to finish executing, since the Step Function runs asynchronously.) Now the verification Step Function also starts with a {\tt Map} workflow, running verifications on each contest in parallel. For each contest, a {\tt Parallel} workflow is invoked, with the signature verification and ZKP verification being processed in parallel. Upon completion of these verifications, the Lambda function that processes results is invoked. If these verifications pass, then this Lambda function stores the encrypted contests and sends them to the SQS queue for the totals to be updated. We stress that with this routing, no voter information enters the Step Function. The ballot status is updated based on the random AnonID and voter information is only handled by the ballot submission Lambda function.

To test the effectiveness of this parallelization on ballot submission, we ran two sets of experiments on ballots of 1 to 5 contests. The first set compared serial versus parallel running times of the digital signature lambda function. The second recorded running times of the full ballot submission process -- both responsiveness on the frontend and the various backend lambda functions. In each experiment, we submitted 300 ballots after a ``warm-up" phase that was designed to invoke several concurrent lambda functions thus reducing the overhead incurred by invoking lambda functions during the experiment. In this way, we simulated a live election environment with a voting rate high enough to require multiple concurrent lambda function invocations. For the first set of experiments, we submitted five ballots per second and for the second set, we submitted one ballot per second.

Table \ref{tab:blindsig} compares serial versus parallel processing of blind signatures and shows that parallelization is the better processing mode for any ballot with more than one contest. As expected, the average running time does not vary significantly over the length of the ballot for parallel processing. Out of 1200 ballots with more than 1 contest, obtaining blind signatures on the ballot contests takes an average of 1.499 seconds.
\begin{table}[ht]
\centering
\begin{tabular}{ c l c c c }
\toprule
 Contests & Processing & Min.\ (ms) & Max.\ (ms) & Mean (ms) \\
\midrule
                 1  & Serial    & 1095 & 2411 & 1408 \\
\hline
 \multirow{2}{*}{2} & Parallel  & 1146 & 2637 & 1457 \\
                    & Serial    & 1526 & 3280 & 1812 \\
\hline
 \multirow{2}{*}{3} & Parallel  & 1147 & 4826 & 1536 \\
                    & Serial    & 1916 & 3775 & 2281 \\
\hline
 \multirow{2}{*}{4} & Parallel  & 1146 & 2771 & 1494 \\
                    & Serial    & 2365 & 5619 & 2779 \\
\hline
 \multirow{2}{*}{5} & Parallel  & 1143 & 2789 & 1508 \\
                    & Serial    & 2759 & 4402 & 3050 \\
\bottomrule
\end{tabular}
\caption{Blind signature Lambda function running times.}
\label{tab:blindsig}
\end{table}

The second set of experiments parallelized the entire ballot submission process. For each ballot submission in this case, we generated a random encrypted ballot and its ZKP on the frontend, obtained a blind signature, removed the masking factor on the blind signature, and submitted the ballot package to the backend. Table \ref{tab:frontend} gives results on the clock time experienced on the frontend to the receipt of the AnonID. A best-fit line of the median times indicates that each contest requires 336ms of computation time.
\begin{table}[ht]
\centering
\begin{tabular}{ c r r r }
\toprule
 Contests & Min.\ (ms) & Max.\ (ms) & Med.\ (ms) \\
\midrule
1 & 2968 & 10524 & 3753 \\
2 & 3102 & 10958 & 4147 \\
3 & 3556 &  9057 & 4493 \\
4 & 3484 & 15801 & 4676 \\
5 & 3767 & 11210 & 5168 \\
\bottomrule
\end{tabular}
\caption{Ballot submission running times on the frontend with parallelized backend.}
\label{tab:frontend}
\end{table}
In a live election, the frontend encrypts a contest and generates its ZKP immediately after a voter makes a selection in that contest, that is, before a voter submits his ballot. So in that case, most of the 336ms per-contest computing time will have been completed before ballot submission. With this new architecture, therefore, most voters will wait between about 3 and 15 seconds from ballot submission to receive their AnonID.

In what follows, we break down the backend operations. Table \ref{tab:backend} lists the running times of the six Lambda functions used in ballot submission: (1) Obtain a blind signature, (2) Overall ballot submission, (3) Verify the blind signature on a contest, (4) Verify the ZKP on a contest, (5) Process the signature and ZKP verifications, and (6) Update the Paillier-encrypted ballot totals. When ballot submission is run in parallel on the backend, the Process Verifications Lambda function is the only function whose average running time varies as a function of the number of contests on a ballot.
\begin{table}[ht]
\centering
\begin{tabular}{ l r r r }
\toprule
 Step & Min.\ (ms) & Max.\ (ms) & Mean (ms) \\
\midrule
Blind Signature         & 1085 & 6212 & 1588 \\
Submit Ballot           & 1384 & 7646 & 1942 \\
Verify Signature        &  375 & 1381 &  437 \\
Verify ZKP              & 2410 & 4146 & 2529 \\
Process Verifications   &  390 & 3651 & 1365 \\
Update Totals           &  228 & 2632 &  341 \\
\bottomrule
\end{tabular}
\caption{Lambda function running times with parallelized backend.}
\label{tab:backend}
\end{table}
From this, we can obtain a few reasonable estimates. First, we can estimate how long the frontend would have to wait to receive the AnonID if the backend processed all verifications in parallel before returning the AnonID. Second, we can estimate how long the frontend would have to wait to receive the AnonID if the backend ran all computations in serial before returning the AnonID. Finally, if everything runs in parallel, we can estimate how long voters need to wait before checking if the ballot with their AnonID has passed verifications.

The Step Function that the Submit Ballot Lambda function calls runs the Verify Signature and Verify ZKP Lambda functions in parallel and the Process Verification Lambda function collects and processes these verifications, calls an API to record the ballot's AnonID on the blockchain, and sends each encrypted contest to an SQS queue for totaling. This results in a variable running time for the Process Verifications function and the Step Function. Tables \ref{tab:processresults} and \ref{tab:stepfunction} list the respective running times, based on the number of contests in the ballot.

\begin{table}[ht]
\centering
\begin{tabular}{ c r r r }
\toprule
 Contests & Min.\ (ms) & Max.\ (ms) & Mean (ms) \\
\midrule
1 &  390 & 2045 &  532 \\
2 &  676 & 2780 &  908 \\
3 &  904 & 3416 & 1171 \\
4 & 1050 & 3620 & 1466 \\
5 & 1377 & 3651 & 1749 \\
\bottomrule
\end{tabular}
\caption{Running times for the Process Verification Lambda function.}
\label{tab:processresults}
\end{table}
\begin{table}[ht]
\centering
\begin{tabular}{ cccc }
\toprule
 Contests & Min.\ (ms) & Max.\ (ms) & Mean (ms) \\
\midrule
1 & 2789 & 4604 & 3066 \\
2 & 3108 & 5695 & 3474 \\
3 & 3396 & 4774 & 3743 \\
4 & 3532 & 6913 & 4061 \\
5 & 3864 & 7182 & 4331 \\
\bottomrule
\end{tabular}
\caption{Running times for the Ballot Verification Step Function}
\label{tab:stepfunction}
\end{table}

Combining the results from Tables \ref{tab:blindsig}-\ref{tab:stepfunction}, we compare average elapsed times for: (a) a fully parallelized ballot submission process, with the AnonID returned before verifications, (b) a fully parallelized ballot submission process, with the AnonID returned after all verifications, (c) all a fully serialized ballot submission process, and (d) the amount of time for a voter to wait before checking an AnonID for completion of the associated ballot verifications.
\begin{table}[ht]
\centering
\begin{tabular}{ c r r r r}
\toprule
 Contests & (a) Parallel & (b) Par.\ w/ AnonID & (c) Serial & (d) Check AnonID \\
\midrule
1 & 3753 & 6819 &  6819 & 3066\\
2 & 4147 & 7621 &  9720 & 3474\\
3 & 4493 & 8236 & 12981 & 3743\\
4 & 4676 & 8737 & 16303 & 4061\\
5 & 5168 & 9449 & 19386 & 4331\\
\bottomrule
\end{tabular}
\caption{Mean running times for serial vs.\ parallel backend}
\label{tab:serialvsparallel}
\end{table}

For ballots with more than 5 contests, it is apparent that serialization would present a significant bottleneck and a poor user experience; a best-fit line of the data in column (c) indicates a per-contest processing time of about 3.17s. A fully parallelized approach, on the other hand, has a 0.336s per-contest processing time, most of which is performed before submitting the ballot. The time to process ballot verifications in that case -- the time a user must wait before checking on the status of the AnonID -- is about 2.8 seconds, plus 321ms per-contest. Both are reasonable wait times from a voter's perspective.

\section{Considerations for Scaling}
\label{scaling}
The number of voters using the Voatz platform in public elections has increased from under 20 voters in the 2018 Primaries\cite{nass2019} to the point where we could handle 100 million. In this section, we discuss the various computing constraints of our AWS backend, estimate the maximum number of voters our architecture could support without modifications, and what changes could be made to increase that limit. We will focus on the two most significant computing services which permit us to handle elections at a large scale: Lambda functions and SQS queues. We note that improvements could certainly be made on the algorithmic level as well, such as using more efficient cryptographic protocols, for example. Such important study is reserved for future work and is outside the scope of this article.

The two most significant bottlenecks in the overall architecture of a homomorphic voting application are the allowable number of concurrent Lambda function invocations and the process of updating the encrypted totals. To that end, we need to know the average running time of the Lambda function that updates totals, the quota on the number of concurrent Lambda function invocations, the throughput of the SQS queue, the maximum length of a queue, and the maximum wait time of a queue. Another factor to consider when discussing these limitations is voting rate. The rate at which people vote fluctuates and is generally the highest during the first and last couple hours of the voting period. Therefore, we will estimate points at which one needs to introduce adjustments to the architecture to support larger elections based on those hours with the highest voting rates.

% Diagram of voting rates?

First, ``tens of thousands" of Lambda functions can be invoked concurrently in a given region (i.e., spanning all API calls), with a default of 1000 concurrent executions \cite{awslambda}. In the Mexican elections, the number of concurrent Lambda function invocations peaked at 39 in the final hour of voting. This included all activity from both the remote and in-person voting modalities. Since this metric scales linearly with the number of voters, we estimate that our architecture could process up to $144734\cdot1000/39 \approx 3.7$ million voters with this default limitation. This limit could be increased by a factor of 10 to 100 by increasing concurrency quotas, and even higher by deploying clones of the web app to different AWS regions. The architecture is sufficient, therefore, to handle elections on a national scale, so we need not be concerned with the constraint on the number of allowable concurrent Lambda function executions.

Shifting our focus to updating totals, we consider SQS queues. SQS queues have a very high throughput -- up to 70,000 messages per second, with a default of 300 messages per second. The length of time that messages can remain in a queue is configurable between 1 minute and 14 days, with a default of 4 days. For the sake of timely results reporting, though, it would be desirable for all queues to have finished processing ballots within an hour after the close of the election. There is no limit on the number of queues nor on the number of messages in a queue in a given region \cite{awssqs}. In the Mexican election, the highest voting rates were in the first hour of remote voting and the last hour of in-person voting: 1593 and 2572 ballots, respectively. This translates to 4632 and 7281 encrypted contests in those two hours (and therefore, SQS messages), respectively. So at the peak, SQS was processing roughly 2 contests per second. Based on the default SQS throughput, the election could support roughly $122496\cdot 300\cdot 3600/4632 \approx 28.6$ million remote and $22238\cdot 300\cdot 3600/7281 \approx 3.3$ million in-person voters voting in a similar time span with a similar ballot. To scale to a larger election, we would simply create more queues, provided that each contest total could only be updated by a Lambda function queued up by a single queue. This approach was in fact used in the Mexican election. We used two queues -- one for remote voting and one for in-person voting. Each contest had separate subtotals for each state and voting modality. In the case of the race for President, for example, there were therefore 64 subtotals. (The motivation for this approach, however, was the desire to generate results reports for each state and modality as quickly as possible after the election closed.)

Finally, we estimate the voting volume that would lead to a 1-hour backlog in an SQS queue. The Lambda function that updated totals had an average running time of 174 ms, so if the voting rate remains under roughly 5.75 ballots per second for each contest, then the SQS queue will essentially remain empty. A 1-hour wait time in the queue would therefore imply roughly $3600/0.174 \approx 20700$ ballots from a single jurisdiction that could not be processed during the election window. In the case of Mexico, 45773 of the 184326 expatriate voters\footnote{39,592 ex-patriate voters voted using a non-internet modality.}, or roughly a quarter, were from the Federal District of Mexico City \cite{ine}. This would translate to roughly 83300 total voters with backlogged ballots in each modality. Now with the aforementioned voting volume of 1593 and 2572 ballots in the first hour of remote voting and the last hour of in-person voting, respectively, we expect that the queues would have been sufficient to process roughly $122496 \cdot 3600/(1593 \cdot 0.174) \approx 1.6$ million remote voters and $22238 \cdot 3600/(2572 \cdot 0.174) \approx 180000$ in-person voters per contest (i.e., voters from the largest state) voting in a similar time frame (2 weeks for remote and 16 hours for in-person voting), without a backlog. Considering the proportion of voters registered in Mexico City, this translates to roughly 6.4 million remote and 720,000 in-person voters. Adding in the extra 83,300 voters, the current architecture could reliably process roughly 6.5 million remote voters over a 2-week span and about 800,000 in-person voters over a 16-hour span. Scaling this to a larger election, therefore, one would simply create additional SQS queues to handle larger voting volumes, with dedicated queues at the precinct, county, state, or regional levels, for example.

This analysis shows that the most significant bottlenecks for such an election are the number of concurrent Lambda function invocations and the efficiency of the Lambda function that updates the totals for a particular contest. If the expected voting volume is expected to exceed the capacity of the SQS queue and the Lambda function that updates these totals, the straightforward mitigations are to simply increase the quota on concurrent Lambda function executions, create more queues, and divide encrypted totals into multiple subtotals. Therefore, the queueing architecture we used could support a much larger election and can readily scale to a national level, up to 100 million or more voters.

\section{Conclusions}
\label{conclusions}

In this article, we described how we secured the voting and reporting processes of the expatriate component of the 2024 Mexican elections. Homomorphic encryption, with the decryption key split using a threshold scheme, prevents election results from being known and released until after the close of the election. Furthermore, it allows for a timely reporting of results. All election totals were decrypted within seconds and full results, with write-in candidates, were generated within a few minutes after election officials downloaded all individually encrypted ballots. Tallies from the individual ballots matched the encrypted running tallies and the number of ballots cast matched the number of voters who voted. Zero-Knowledge Proofs and blind digital signatures allowed voters to cast their ballots anonymously and verify that their ballots were unaltered. Various measures prevented a voter from voting more than once and ensured that each ballot was tallied exactly once. Our security measures assumed that an adversary had access to source code and the ability to read, copy, and modify all traffic between the frontend and backend. All this lends confidence to the notion that internet voting, when designed with thorough testing and thoughtful risk mitigation for security, accessibility, availability, and scalability, is a safe and viable means to conduct a public election.

Our experiments and analysis showed how to overcome certain bottlenecks. We showed how to parallelize ballot submission efficiently using AWS Step Functions in order to accommodate longer ballots and to improve the voter experience. We also analyzed the capabilities of various AWS computing services, such as Lambda functions and SQS queues, to show that a secure election could be run on a national scale over the internet.

We claim, therefore, that voting via the internet using techniques described in this paper is a feasible and secure solution for public elections. In the United States, internet voting is currently restricted to voters registered in a handful of states who cannot make it to a polling location in their home precinct, including voters with disabilities, overseas voters, and deployed military personnel \cite{ncsl}. We would also obviously want robust voter identification methods to verify that the person voting is a registered citizen, such as those used in the Voatz Mobile App and the KYC/AML identification techniques in common use by banks and other institutions. As a first step towards implementing such a solution in a United States election, though, some states could allow in-person voting at selected American embassies and consulates, with voters voting on a kiosk much like what was done for Mexico. Voting can progress beyond the outdated and error-prone methods in common use today to enable provable results efficiently and restore provable confidence in the election process.

\section*{Acknowledgements}
With sincere gratitude, we thank INE (Mexico's National Electoral Institute) and AWS for their support and partnership in this project. We also extend our sincere gratitude to an anonymous reviewer who gave us very helpful feedback and posed many insightful questions on our initial submission.

\bibliographystyle{plain}
\bibliography{mexico}

\end{document}